\documentclass[conference]{IEEEtran}
\IEEEoverridecommandlockouts
\usepackage{cite}
\usepackage{amsmath,amssymb,amsfonts}
\usepackage{algorithmic}
\usepackage{graphicx}
\usepackage{textcomp}
\usepackage{xcolor}
\usepackage{hyperref}
\usepackage{wrapfig,lipsum,booktabs}

\def\BibTeX{{\rm B\kern-.05em{\sc i\kern-.025em b}\kern-.08em
    T\kern-.1667em\lower.7ex\hbox{E}\kern-.125emX}}
\begin{document}

\title{Improving Homograph Attack Classification\\
}

\makeatletter
\newcommand{\linebreakand}{%
  \end{@IEEEauthorhalign}
  \hfill\mbox{}\par
  \mbox{}\hfill\begin{@IEEEauthorhalign}
}
\makeatother

\author{\IEEEauthorblockN{Tran Phuong Thao}
\IEEEauthorblockA{%\textit{dept. name of organization (of Aff.)} \\
\textit{The University of Tokyo}\\
Tokyo, Japan \\
tpthao@yamagula.ic.i.u-tokyo.ac.jp}
}

\maketitle

\begin{abstract}
A visual homograph attack is a way that the attacker deceives the web users about which domain they are visiting by exploiting forged domains that look similar to the genuine domains. T. Thao et al.~\cite{IFIPSec19} proposed a homograph classification by applying conventional supervised learning algorithms on the features extracted from a single-character-based Structural Similarity Index (SSIM). This paper aims to improve the classification accuracy by combining their SSIM features with 199 features extracted from a N-gram model and applying advanced ensemble learning algorithms. The experimental result showed that our proposed method could enhance even 1.81\% of accuracy and reduce 2.15\% of false-positive rate. Furthermore, existing work applied machine learning on some features without being able to explain why applying it can improve the accuracy. Even though the accuracy could be improved, understanding the ground-truth is also crucial. Therefore, in this paper, we conducted an error empirical analysis and could obtain several findings behind our proposed approach. 

\end{abstract}


\section{Introduction}
\label{section:introduction}
Visual homograph attack is a type of spoofing attack in which the attackers deceive the users about what domains they are accessing by using the fake domains that look like the genuine domains. 
E. Gabrilovic~\cite{Gabrilovic2002} first introduced the attack using a fake domain \emph{microsoft}, which incorporates the Russian letters ‘c’ (U+0441) and ‘o’ (U+043E). 
Then, in 2017, Z. Xudong~\cite{Xudong} demonstrated the seriousness of the attack by registering the homograph \emph{apple.com} (targeting the brand domain of Apple Inc.), which replaces the regular Latin letters ‘a’, ‘p’, ‘l’, and ‘e’ by the Cyrillic letters ‘a’ (U+0430), ‘p’ (U+0440), ‘l’ (U+04CF), and ‘e’ (U+0435). The demonstration attracted a lot of interest from media~\cite{Kieren, LeeM, MohitK, AlexHern}. A large-scale analysis~\cite{Reexamination2018} on International Domain Names (IDNs) in 2018 showed that, just for the top 1,000 Alexa brand domains, more than 1,516 IDN homographs were registered. Not just growing with the large number, the attack also becomes more progressive and sophisticated.

Several homograph detections have been proposed. While most of the papers focus on IDNs, a state-of-the-art paper~\cite{IFIPSec19} can thoroughly deal with the homographs not just in IDNs but also in English domains. Instead of determining the homographs by picking the domains with visual similarity scores greater than a fixed threshold, the authors proposed a machine learning-based classification using the visual similarity as features to address the high false-positive rate caused by the fixed similarity threshold. Unlike other papers that calculated the visual similarity on the entire domain string, their approach is based on the similarity calculated on every character in the domain string, which can improve the classification accuracy to even 8.62\% from the previous approaches. 

In this paper, we found that there is room for improvement. First, we observed from the result in~\cite{IFIPSec19} that decision tree gives the best performance compared with other conventional classification algorithms (i.e., support vector machine, naive bayes, nearest neighbors, etc.). So if the decision tree could be improved, an even better result might be obtained. Therefore, we applied advanced decision tree-based algorithms known as \emph{ensemble learnings}, which combine the multiple decision trees with seeing if the overall performance can be improved. Second, while~\cite{IFIPSec19} utilized the structural similarity index (SSIM), which is a modern visual similarity measure in image processing, we question whether a similarity measure used in text processing can give some extra information. Therefore, in this paper, we tried to combine the SSIM with 199 text features extracted from the N-gram model. Our experimental result showed that the accuracy could be increased up to 1.81\% and the false-positive rate can be reduced even 2.15\%. Last but not least, most of the previous work just applied machine learning without being able to explain the reason behind why it can improve the accuracy. Since the ground-truth of the proposed method is also essential, we conducted an empirical \emph{error analysis} technique by Andrew Ng~\cite{erroranalysis}. We found that the SSIM performs worse than the unigram when the SSIM is lower than 0.959, performs equally to the unigram when the score lies within the range $[0.959, 0.989]$, but performs significantly better than the unigram when the score is greater than 0.989. The combination approach outperforms both the SSIM and unigram in any case. While the unigram can detect the abnormal samples with a low SSIM score but are homographs better than the SSIM, the SSIM can detect the abnormal samples with a high SSIM score but are non-homographs much better than the unigram. 

The rest of this paper is organized as follows. The related work is introduced in Section~\ref{section:relatedwork}. The proposed approach is presented in Section~\ref{section:proposedapproach}. The experiment is given in Section~\ref{section:experiement}. The discussion is described in Section~\ref{section:discussion}. Finally, the conclusion is drawn in Section~\ref{section:conclusion}.

\section{Related Work}
\label{section:relatedwork}
Homograph detections can be categorized as follows.

\subsection{Disabling Automatic IDN Conversion}
Web browsers disabled the feature of automatic IDN conversion such as Google Chrome~\cite{chromefix}, Safari~\cite{safarifix}, Mozilla Firefox~\cite{Firefoxfix}, Internet Explorer~\cite{IEfix}.
%Opera~\cite{opera2007}. 
Instead of displaying non-ASCII characters (e.g., $g\tilde{o}\tilde{o}gle.com$), the browsers now display the Punycode form (e.g., \emph{xn\texttt{-{}-}ggle-0qaa.com}). The method is simple but has some drawbacks. First, the homographs exploit not just the international characters but also the look-alike characters in ASCII itself, and thus disabling the automatic IDN conversion is ineffective. For example, the Punycode form of \emph{bl0gspot.com} (targeting \emph{blogspot.com}) is still \emph{bl0gspot.com}. Second, when a user accidentally accesses a homograph, the browser displays its Punycode on the address bar. Even if a warning message is shown, it is a burden when they frequently access IDNs. After clicking/typing the homograph, the user often concentrates on the content rather than the address bar. Also, due to a large number of international users and over 9 million registered IDNs worldwide (as of 2020~\cite{numberIDN2020}), disabling the IDN conversion is not a convenient solution. 

\subsection{Homograph Generation Tools}
\label{relatedwork:tools}
Several tools were developed to help generate homograph IDNs~\cite{Dnstwist, TimoF, AlissonM, Domaininsta, DNPedia, AdrianC}.
%~\cite{Dnstwist, TimoF, AlissonM, RemcoV, Domaininsta, DNPedia, AdrianC}. 
They defined a set of mistakable keystrokes on a keyboard (the three common keyboards are qwerty, qwertz, and azerty) and a set of visually confusable characters. On inputting a domain, the algorithms are built for bit-squatting, insertion, omission, repetition, replacement, subdomain, hyphenation, transposition, vowel swapping, and addition. There are some shortcomings. First, the sets (i.e., mistakable keystrokes and visually confusable characters) were defined in a subjective perspective of the developers with no standards. Seconds, the defined sets are very limited because getting all the large set's permutations will take a heavy computational cost. Therefore, the generated homographs is insufficient.

\subsection{Large-scale Domain Registration Databases}
Analyses were conducted to find homographs using auxiliary databases containing registered domains. L. Baojun et al.~\cite{Reexamination2018} scanned two top-level domains (TLDs) (including gTLDs (generic TLDs) and iTLDs (international TLDs)) and compiled 1.4 million IDNs. The top 1,000 domains in Alexa ranking were extracted. The image of each IDN was compared to that of each brand domain using the Structural Similarity (SSIM). They defined a threshold for SSIM as 0.95, picked up the IDNs with SSIM over or equal to 0.95, and defined them as homographs. C. Daiki et al.~\cite{RAID19} leveraged the commercial Whois database to extract 4.4 million IDNs. 2,310 top brand domains in Alexa, Umbrella, and Majestic were extracted. Similar to~\cite{Reexamination2018}, they calculated SSIM but chose a threshold as 0.99 instead of 0.95. The domains whose SSIMs 0.99 are chosen as homographs. Two main weak points exist. 
%First, they defined the fixed high SSIM thresholds (i.e., 0.95 and 0.99) to reduce the false-positive rate. However, 
First, non-homographs still can have an SSIM higher than these thresholds. For example, \emph{so.com} and \emph{sp.com} are not related together at all in either content or visual appearance with most people, but the SSIM is very high (0.997). Vice versa, some homographs have low SSIMs. For example, $b\hat{i}ttr\hat{e}x.com$ is a homograph targeting \emph{bittrex.com} but the SSIM is 0.916 which is under their thresholds. Second, the methods may lack the scalability since every day there are many newly registered domains (e.g., just in 2020/06, there were 3,271,015 new domains were registered~\cite{newdomains2020}).
%(average over 100,000 new domains per day). 

\subsection{Machine Learning-based Classification}
K. Tian et al.~\cite{TianK} generated homographs using typo-squatting, bits-squatting, glyph-squatting, combo-squatting, and wrong-TLD squatting. They applied visual analysis and optical character recognition (OCR) to extract the page screenshots' key visual features.  A. Pieter et al.~\cite{Pieter2015} focused on typo-squatting homographs only generated using missing-dot typos, omission typos, permutation typos (consecutive characters are swapped), substitution typos, and duplication typos. A Whois lookup was performed for each domain. They visited each page of the domains and measured its visual appearance to construct a cluster using the concatenation of a perceptual hash of the page’s screenshot and a locality-sensitive hash of its HTML body. T. Thao et al.~\cite{IFIPSec19} claimed that a homograph domain is a type of spoofing attack but does not need to have malicious or phishing content. When the attacker registers a look-alike domain targeting the brand domain, it is already misbehavior even though the page content is blank. How to \emph{proactively} detect the homographs right after it is registered is even more crucial than just responding to it after it has happened. Furthermore, unlike other previous papers which only address homographs belonging to IDNs, ~\cite{IFIPSec19} pointed out that homographs do not need to be IDNs since the attack can exploit also look-alike Latin characters such as (‘1’, ‘l’, ‘i’), (‘o’, ‘0’), (‘rn’, ‘m’). The authors, thus, proposed a method to deal with homographs both in IDNs and non-IDNs. Moreover, other papers chose a high visual similarity (i.e., SSIM) threshold and determined the homograph by picking the IDNs with SSIM scores greater than the threshold. However, in fact, many domains have such a high visual similarity but are non-homographs; and vice versa, many domains have a low visual similarity but are homographs. Moreover, they proved that applying the visual similarity for the whole domain strings can lead to a low accuracy because many domains are too visually different with the brand domains but have high SSIM scores; for example, ``aa.com'' and ``ea.com'' have 0.952 SSIM and was listed as a homograph but is not a homograph. Therefore, the authors in~\cite{IFIPSec19} proposed a machine learning-based classification using SSIM scores calculated on every character in the domain string to improve the accuracy and lower the false-positive rate caused by the fixed similarity threshold and the entire-string-based SSIM.

\section{Proposed Approach}
\label{section:proposedapproach}
\subsection{Data Collection}
The process to collect the data consists of four steps.  

\subsubsection{Generating Homographs}
Homographs were collected from different sources. From the Confusable Unicode (CUni)~\cite{unicodelist} which contains over 6,000 pairs of confusable characters, we generated the permutations by replacing each character in the given brand domains by the confusable characters in CUni. We got 26,021 homograph candidates. Using homograph generation tools~\cite{TimoF, AlissonM, Dnstwist} (see Section~\ref{relatedwork:tools}), we generated another 12,338 candidates. Also, thanks to the authors of~\cite{Reexamination2018} for sharing us 1,516 candidates from matching 1,000 top Alexa domains with 1.4 million registered IDNs. 

\subsubsection{Filtering Active Homographs and Generating Non-homographs}
\label{section:datacollectfilter}
We extracted the unique domains and checked if the domains are registered by querying Whois registration records. Whois contains the information such as registration name, organization, address, creation date, expiration date, etc. We got 1,174 unique domains. 
From each brand domain used to generate 1,174 homograph candidates, we generated at least one domain that strongly looks different from the brand domains to ensure they are non-homographs. 1,969 non-homograph candidates were generated. We got 3,143 samples in total. It may be relatively small, but in fact, the number of homographs is not so many. \cite{Reexamination2018} and~\cite{RAID19} scanned an entire commercial Whois database but only found 1,516 and 2,310 homographs, respectively. 
%If we increase the dataset by generating more non-homographs, it will cause an imbalance distribution.

\subsubsection{Sample Relabeling}
The samples are re-labeled by humans. A brand domain may have multiple homographs, but a homograph can only target one brand domain. Thus, the homographs that have multiple brand domains need to be re-labeled. For example, \emph{wikipedia.org} and \emph{wikimedia.org} look-alike with each other but are two different brand domains.
The volunteers assessed each pair of brand domain and homograph. For example, for the pairs (\emph{wikipedia.org}, \emph{wikinedia.org}) and (\emph{wikimedia.org}, \emph{wikinedia.org}), ``n" and ``m" are closer in the keyboard rather than ``n" and ``p". We, therefore, determine \emph{wikinedia.org} is a homograph of \emph{wikimedia.org} instead of \emph{wikipedia.org}. %The feature (i.e., SSIM) will be calculated between \emph{wikinedia.org} and \emph{wikimedia.org}. 
Humans may have different opinions about the homographs (e.g., A may think \emph{esss.com} is a homograph of \emph{asss.com}, but B may not think so). Therefore, we asked an odd number of volunteers, and the final decision is determined based on the majority vote. 
%The people were asked to re-label the samples when assuming they are actual attackers. 
We finally got 1,073 samples labeled as homographs and 2,070 samples labeled as non-homographs.

\subsection{Feature Extraction}
\subsubsection{Structural Similarity (SSIM)}
\label{section:computeSSIM}
SSIM is the most common method for measuring the visual similarity between two images. SSIM was proposed as a perceptual measure based on visible structures in the images to improve the traditional methods such as Peak Signal-To-Noise Ratio (PSNR) or Mean Squared Error (MSE). For two images $x$ and $y$ with the same size $m \times m$, the SSIM is calculated:
\begin{equation}
    SSIM(x, y) = \frac{(2\mu_x\mu_y+r_1)(2\sigma_{xy}+r_2)}{(\mu^2_x+\mu^2_y+r_1)(\sigma^2_x+\sigma^2_y+r_2)}
\end{equation}
$\mu_x$ and $\mu_y$ denote the average of $x$ and $y$. $\sigma^2_x$ and $\sigma^2_y$ denote the variance of $x$ and $y$. $\sigma_{xy}$ denotes the covariance of $x$ and $y$. $r_1=(k_1L)^2$ and $r_2 = (k_2L)^2$ denote the variables used to stabilize the division with weak denominator. $L=2^{\#\text{bits per pixel}}-1$ denotes the dynamic range of the pixel-values and $k_1 = 0.01$, $k_2 = 0.03$ by default. The SSIM ranges within $[-1, +1]$ where 1 indicates a perfect similarity.

For two domains $D_x$ and $D_y$, if they have different numbers of characters, the SSIM is set to 0.%which indicates that $D_x$ and $D_y$ are completely different. 
Otherwise, let $n$ denote the number of characters: $D_x = \{c_{x1}, \cdots, c_{xn}\}$ and $D_y = \{c_{y1}, \cdots, c_{yn}\}$. A simple method is to transform the entire $D_x$ and $D_y$ to two images $I(D_x)$ and $I(D_y)$ and calculate $SSIM(I(D_x), I(D_y))$. However, it can increase the false-positive rate. For example, $\grave{a}a.com$ and \emph{ea.com} have a high SSIM (0.952) but $\grave{a}a.com$ is not a homograph of \emph{ea.com}. We leverage the idea from~\cite{IFIPSec19} which calculated the SSIM based on each character. $\forall i \in [1, n]$, we calculate $SSIM(I(c_{xi}), I(c_{yi}))$ where $I(c_{xi})$ and $I(c_{yi})$ denote the images of $c_{xi} \in D_x$ and $c_{yi} \in D_y$. The SSIM between $D_x$ and $D_y$ is calculated as the average SSIM:
\begin{equation}
SSIM(D_x, D_y) = \frac{\sum_{i=1}^{n} SSIM(I(c_{xi}), I(c_{yi}))}{n}  
\label{eq:SSIM}
\end{equation}
This average SSIM is used as the feature for the model. 

\subsubsection{Unigram}
Let $N$ denote the number of domain samples from the dataset ($N=3,143$). Each domain contains a different set of characters: $\{D_{j}=\{c_{j1}, \cdots, c_{jn_{j}}\}\}$ where $j \in [1, N]$ and $n_{j}$ denotes the number of characters in each set. 
%$D_{1}=\{c_{11}, \cdots, c_{1n_{1}}\}$, $\cdots$, $D_{N}=\{c_{N1}, \cdots, c_{1n_{N}}\}$ where $n_{1}, \cdots, n_{N}$ denote the number of characters in each set. 
The vector of unique characters in all the samples is:
\begin{equation}
V_{unigram} = \cup^{N}_{i=1} D_{i}
\end{equation}
Each element in $V_{unigram}$ is used as a unigram feature. Suppose $V_{unigram}$ consists of $\lambda$ unique characters: $V_{unigram} = \{c_{v1}, \cdots, c_{v\lambda}\}$ corresponding to $\lambda$ unigram features $\{f_{v1},\cdots, f_{v\lambda}\}$. For each $f_{vi}$ ($i \in [1, \lambda]$), the unigram feature for $D_{j}$ ($j \in [1, N]$) is calculated:
\begin{equation}
f_{vi}(D_{j}) = 
\begin{cases}
1 & \text{if} \quad c_{vi}\in D_{j} \\
0 & \text{if} \quad c_{vi}\not \in D_{j}
\end{cases}
\end{equation}
The unigram vector for $D_{j}$ is constructed: $F_{unigram}(D_j) = \{f_{v1}(D_{j})$, $\cdots$, $f_{v\lambda}(D_{j})\}$. From 3,143 samples, $\lambda=199$ unigram features were finally extracted.

\subsection{Training}

Ensemble algorithms combine base estimators to produce one optimal predictive estimator with better performance using two different strategies: boosting and averaging. The boosting strategy builds the base estimators sequentially. Each base estimator is used to correct and reduce the bias of its predecessor. Two common boosting algorithms include:
\begin{itemize}
    \item AdaBoost: the sample distribution is adapted to put a higher weight on the samples that are misclassified and a lower weight on the correctly classified samples. 
    \item GradientBoost: instead of weighting the samples, it trains the negative gradient of the loss function and builds the subsequent learners to predict the loss (the difference between the predicted value and the real value). 
\end{itemize}

The average strategy builds the estimators independently. The predictions are averaged based on the aggregated results. 
%The combined estimator reduces the variance. 
Three common average algorithms include:
\begin{itemize}
\item RandomForest: implements a meta estimator that fits several decision tree classifiers on various randomized sub-samples of the dataset and uses averaging to create the best predictive estimator. A bootstrap is created by randomly sampling the dataset with a replacement. The sub-samples' size is set to be the same as the size of the original input sample. The decision tree is trained by recursively splitting the data (converting the non-homogeneous parent into the two most homogeneous child nodes). 
RandomForest selects an optimal split on the features at every node. 
\item ExtraTrees: also produces the best estimator but has some differences. While RandomForest uses the optimal split and sets $bootstrap = True$, ExtraTrees uses the random split and sets $bootstrap = False$. RandomForest supports drawing sampling with a replacement, but ExtraTrees supports it without a replacement.
\item Bagging: While RandomForest and ExtraTrees select only a subset of randomized features for splitting a node, Bagging uses all the features for splitting a node.
\end{itemize}

We use $k$-fold cross-validation and measure the accuracy $ACC = \frac{tp+tn}{tp+tn+fp+fn}$, false-positive rate $FPR = \frac{fp}{fp+tn}$, and true positive rate $TPR = \frac{tp}{tp+fn}$ where $tp$, $tn$, $fp$, and $fn$ denote the true positive, true negative, false positive, and false negative values, respectively.

\section{Experiment}
\label{section:experiement}
The program is written in Python 3.7.4 on MacBook Pro 2.8 GHz Intel Core i7, RAM 16GB. The machine learning algorithms are executed using scikit-learn\footnote{Scikit-learn: https://scikit-learn.org/stable/\#} 0.22. Seven algorithms in~\cite{IFIPSec19} were implemented including SVC (support vector machine), NuSVC (nu-libsvm SVC), LinearSVC (linear kernel SVC), GaussianNB (Gaussian naive bayes), MultinomialNB (multinomially distributed naive bayes),  BernoulliNB (Bernoulli distribution naive bayes), NearestCentroid (centroid nearest neighbors), KNeighbors (k-nearest neighbors), DecisionTree (decision tree), MLP (multi-layer perceptron back-propagation-based neural network), and SGD (stochastic gradient descent). For KNeighbors, the number of neighbors is set to 5. For the ensemble algorithms, the number of base estimators is set to 100. $k$-fold in cross-validation is set to 5. 

Table~\ref{table:result} describes the results. $S_{P}$, $U_{P}$, and $SU_{P}$ denote the approach using SSIM, the approach using the unigram, and the combination approach. $SU_{P}$ outperforms $S_{P}$ with 9 over 16 algorithms. Interestingly, using any of the ensemble algorithms, the accuracy of $SU_{P}$ can be increased even more than 1.5\% from that of $S_{P}$. The blue texts represent the results using~\cite{IFIPSec19} in which the best performance is KNeighbors (95.10\% of accuracy, 05.60\% of false-positive rate). The red texts represent the result using our approach in which the best performance is Bagging (96.91\% of accuracy, 3.45\% of false-positive rate). It indicates that our approach could improve 1.81\% of accuracy and reduce 2.15\% of false-positive rate. 

\begin{table*}[!ht]
\centering
\caption{Results of SSIM Approach and Combination Approach}
\begin{tabular}{c l | l l l | c | c | c | c | c | c}
\textbf{No} & \textbf{Algorithm} & \multicolumn{3}{c|}{\textbf{ACC (\%)}} & \multicolumn{3}{c|}{\textbf{FPR (\%)}}   & \multicolumn{3}{c}{\textbf{TPR (\%)}}    \\
& & \textbf{S\textsubscript{P}} & \textbf{U\textsubscript{P}} & \textbf{SU\textsubscript{P}} & \textbf{S\textsubscript{P}} & \textbf{U\textsubscript{P}} & \textbf{SU\textsubscript{P}} & \textbf{S\textsubscript{P}} & \textbf{U\textsubscript{P}} & \textbf{SU\textsubscript{P}}\\
\hline
1 & SVC & \color{blue} \textbf{94.52} & 66.85 & 66.88 & 05.20 & 48.53 & 48.49 & 94.80 & 51.47 & 51.51 \\
2 & NuSVC & \color{blue} \textbf{94.43} & 87.21 & 88.86 & 05.27 & 17.39 & 14.97 & 94.73 & 82.61 & 85.03 \\
3 & LinearSVC & \color{blue} \textbf{94.62} & 94.11 & 94.85 & 05.39 & 07.03 & 05.94 & 94.61 & 92.97 & 94.06 \\
4 & GaussianNB & \color{blue} \textbf{94.53} & 81.71 & 81.71 & 05.20 & 26.22 & 26.22 & 94.80 & 73.78 & 73.78 \\
5 & MultinomialNB & \color{blue} \textbf{65.86} & 91.82 & 91.86 & 50.00 & 10.39 & 10.35 & 50.00 & 89.61 & 89.65 \\
6 & BernoulliNB & \color{blue} \textbf{65.86} & 91.51 & 91.51 & 50.00 & 10.77 & 10.77 & 50.00 & 89.23 & 89.23 \\
7 & NearestCentroid & \color{blue} \textbf{93.19} & 70.95 & 71.59 & 06.08 & 30.17 & 29.61 & 93.92 & 69.83 & 70.39\\
8 & KNeighbors & \color{blue} \textbf{95.10} (*)& 85.78 & 91.19 & 05.60 & 19.60 & 11.73 & 94.40 & 80.40 & 88.27 \\
9 & DecisionTree & \color{blue} \textbf{95.00} & 88.93 & 96.12 & 05.59 & 14.03 & 04.34 & 94.41 & 85.97 & 95.66 \\
10 & MLP & \color{blue} \textbf{94.62} & 94.72 & 94.46 & 05.77 & 06.36 & 06.70 & 94.23 & 93.64 & 93.30 \\
11 & SGD & \color{blue} \textbf{93.92} & 94.30 & 94.85 & 05.65 & 06.71 & 06.08 & 94.35 & 93.29 & 93.92 \\
\hline
12 & AdaBoost & 95.20 & 91.09 & \color{red} \textbf{96.82} & 05.93 & 11.65 & 03.75 & 94.07 & 88.35 & 96.25\\
13 & GradientBoost & 95.04 & 84.98 & \color{red} \textbf{96.63} & 05.76 & 21.04 & 03.70 & 94.24 & 78.96 & 96.30 \\
14 & ExtraTrees & 94.91 & 91.50 & \color{red} \textbf{96.85} & 05.66 & 11.06 & 03.35 & 94.34 & 88.94 & 96.65 \\
15 & RandomForest & 95.00 & 90.93 & \color{red} \textbf{96.88} & 05.59 & 12.07 & 03.37 & 94.41 & 87.93 & 96.62 \\
16 & Bagging & 95.04 & 91.16 & \color{red} \textbf{96.91} (**) & 05.57 & 11.50 & 03.45 & 94.43 & 88.50 & 96.55 \\
\end{tabular}\\
(*): best performance by~\cite{IFIPSec19}. (**): best performance by our approach
\label{table:result}
\end{table*}

\section{Discussion}
\label{section:discussion}

\subsection{Error Analysis}
%This section analyzes why $SU_{P}$ outperforms $S_{P}$. 
We first analyze different SSIM ranges.
The best algorithms were selected: AdaBoost for $S_{P}$, MLP for $U_{P}$, and Bagging for $SU_{P}$. 
%Although the algorithms are different, the experiments were done in the same configuration. 
The SSIM of 3,143 samples range from [0.855, 1.000]. The range is first divided into 10 equal bins (see Table~\ref{table:compare1}). 
%The last three columns present the number of samples correctly predicted by each approach. 
Since the number of samples in each bin is too different, e.g., the first bin only has 3 samples, so the big difference between 33.33\% (for $S_{P}$) and 100\% (for $U_{P}$ and $SU_{P}$) cannot correctly reflect the result. Therefore, different subsets with the same number of samples are considered. 3,143 samples are sorted in ascending order of the SSIM and divided into 10 equal subsets with the same number of samples. For a fair comparison, the same algorithm Bagging is chosen (even if AdaBoost, which is the best algorithm for $S_{P}$ is chosen, $SU_{P}$ still performs better (see Table~\ref{table:result})). 
%We extracted the SSIM sub-ranges for each subset and calculated the number of samples that were correctly predicted by each approach. 
The result is shown in Figure~\ref{fig:errorsample}. $U_{P}$ outperforms $S_{P}$ in the first 6 bins ([0.855, 0.959]), have the same accuracy as $S_{P}$ when the SSIM reaches the 7th bin ([0.959, 0.989]), and is worse than $S_{P}$ when the SSIM lies within the last 3 bins ([0.989, 1.000]). Most importantly, $SU_{P}$ achieves the best accuracy in all the cases (the gray chart is always on top of the blue and orange charts). It indicates that $SU_{P}$ inherits the benefits from both $S_{P}$ and $U_{P}$. Let ``low SSIM range group" (LRG), ``middle SSIM range group" (MRG), and ``high SSIM range group (HRG)" denote the first 6 bins, the 7th bin, and the last 3 bins, respectively.
%Let $R_{low}$, $R_{middle}$, and $R_{high}$ denote the corresponding SSIM ranges in the groups. The notations $R_{low}$ and $R_{high}$ will be used for analyzing error samples below.

\begin{table*}[!ht]
\centering
\caption{Accuracy for Equal SSIM Sub-ranges and for Equal SSIM Sub-ranges}
\begin{tabular}{c c r | r  r | r  r | r  r}
\textbf{No} & \textbf{Bins} & \textbf{\#Samples} & \multicolumn{6}{c}{\textbf{\#Correct Samples (Percentage \%)}} \\
 & & & \multicolumn{2}{c|}{\textbf{$S_{P}$ (AdaBoost)}} & \multicolumn{2}{c|}{\textbf{$U_{P}$ (MLP)}} & \multicolumn{2}{c}{\textbf{$SU_{P}$ (Bagging)}}  \\
 \hline
\multicolumn{9}{c}{Accuracy for Equal SSIM Sub-ranges}\\
1 & [0.855, 0.869]	& 3	& 1 & (33.33\%)	& 3 & (100.00\%) & 3 & (100.00\%)\\
2 & (0.869, 0.884]	& 7	& 7 & (100.00\%) & 5 & (71.43\%) & 5 & (71.43\%)\\
3 & (0.884, 0.898] & 50 & 46 & (92.00\%) & 48 & (96.00\%) & 47 & (94.00\%)\\
4 & (0.898, 0.913] & 282 & 276 & (97.87\%) & 282 & (100.00\%) & 281 & (99.65\%)\\
5 & (0.913, 0.927] & 562 & 556 & (98.93\%) & 558 & (99.29\%) & 560 & (99.64\%)\\
6& (0.927, 0.942] & 558 & 551 & (98.75\%) & 557 & (99.82\%)	& 557 & (99.82\%)\\
7 & (0.942, 0.956] & 371 & 361 & (97.30\%) & 369 & (99.46\%) & 371 & (100.00\%)\\
8 & (0.956, 0.971] & 179 & 161 & (89.94\%) & 175 & (97.77\%) & 176 & (98.32\%)\\
9 & (0.971, 0.985] & 118 & 67 & (56.78\%) & 81 & (68.64\%) & 85 & (72.03\%)\\
10 & (0.985, 1.000] & 1013 & 967 & (95.46\%) & 893 & (88.15\%) & 966 & (95.36\%)\\
\hline
\multicolumn{9}{c}{Accuracy for Equal SSIM Sub-ranges}\\
1 & [0.855, 0.912] & 314 & 297 & (94.59\%) & 307 & (97.77\%) & 309 & (98.41\%)\\
2 & [0.912, 0.920] & 314 & 305 & (97.13\%) & 310 & (98.73\%) & 312 & (99.36\%)\\
3 & [0.920, 0.928] & 314 & 307 & (97.77\%) & 313 & (99.68\%) & 314 & (100.00\%)\\
4 & [0.928, 0.936] & 314 & 312 & (99.36\%) & 312 & (99.36\%) & 314 & (100.00\%)\\
5 & [0.936, 0.945] & 314 & 309 & (98.41\%) & 313 & (99.68\%) & 313 & (99.68\%)\\
6 & [0.945, 0.959] & 314 & 302 & (96.18\%) & 310 & (98.73\%) & 314 & (100.00\%)\\
7 & [0.959, 0.989] & 314 & 236 & (75.16\%) & 236 & (75.16\%) & 250 & (79.62\%)\\
8 & [0.989, 0.995] & 314 & 289 & (92.04\%) & 262 & (83.44\%) & 295 & (93.95\%)\\
9 & [0.995, 0.997] & 314 & 314 & (100.00\%)& 261 & (83.12\%) & 314 & (100.00\%)\\
10& [0.997, 1.000] & 317 & 314 & (99.05\%) & 253 & (79.81\%) & 314 & (99.05\%)
\end{tabular}
\label{table:compare1}
\end{table*}

\subsubsection{Error Samples in LRG}
LRG contains 1,884 samples including 39 homographs (2.07\%) and 1,845 non-homographs (97.93\%) (see Table~\ref{table:lowhigh}). Let error samples denote the samples that are correctly predicted by $U_{P}$ but incorrectly predicted by $S_{P}$. We found 44 error samples in which $SU_{P}$ can correctly detect most of them (43 samples (97.73\%)). The 44 error samples include 17 error homographs (43.59\% of homographs) and 27 error non-homographs (1.46\% of non-homographs). While the percentage of non-homographs is dominant to that of homographs (47.31 times), the percentage of error non-homograph is submissive to that of error homographs (29.86 times). Analyzing the 17 error homographs, we found that most of them have a relatively high ratio (e.g., $\geq 30\%$) of characters that are visually different from those in the brand domains. Formally, let $D_{y}= \{c_{y1}, \cdots, c_{yn}\}$ and $D_{x}= \{c_{x1}, \cdots, c_{xn}\}$ denote an error homograph and its brand domain where $n$ is the number of characters. Let $K_{x} = \{c_{kx1}, \cdots, c_{kx\gamma}\} \subset D_{x}$ and $K_{y} = \{c_{ky1}, \cdots, c_{ky\gamma}\} \subset D_{y}$ denote the subsets that contain different characters such that $SSIM(I(c_{kxi}), I(c_{kyi})) < 1$ where $i \in [1, \gamma]$. In most of $D_{y}$, $K_{y}$ accounts for quite large portion, e.g., $|K_{y}| /n \geq 30\%$. For example, the homograph $\hat{e}k s\ddot{i} s\hat{o}z1\ddot{u}k.com$ and its targeted brand domain \emph{eksisozluk.com}
%$D_{x} = \{e, k, s, i, s, o, z, l, u, k, ., c, o, m\}$ and $D_{y} = \{\hat{e}, k, s, \ddot{i}, s, \hat{o}, z, 1, \ddot{u}, k, ., c, o, m\}$. $D_{x}$ and $D_{y}$ 
have 5 positions with SSIM$<$1, i.e., ($\hat{e}, e$), ($\ddot{i}, i$), ($\hat{o}, o$), ($1, l$), and ($\ddot{u}, u$). Since the overall SSIM is computed based on the average SSIM of all the character pairs, the overall SSIM is reduced when many pairs have SSIM$<$1. 
%Although a low SSIM is not only caused by the number of visually different characters but also by the SSIM score in each character pair, the former affects the entire low SSIM stronger than the latter. 
Usually, the high-SSIM sample tends to be a homograph; and the low-SSIM sample tends to be a non-homograph. Thus, $S_{P}$ detects better than $U_{P}$. However, for the abnormal samples that have a low SSIM but are homographs, $U_{P}$ detects better than $S_{P}$.

\begin{figure}[!ht]
  \centering
 \includegraphics[scale=0.55]{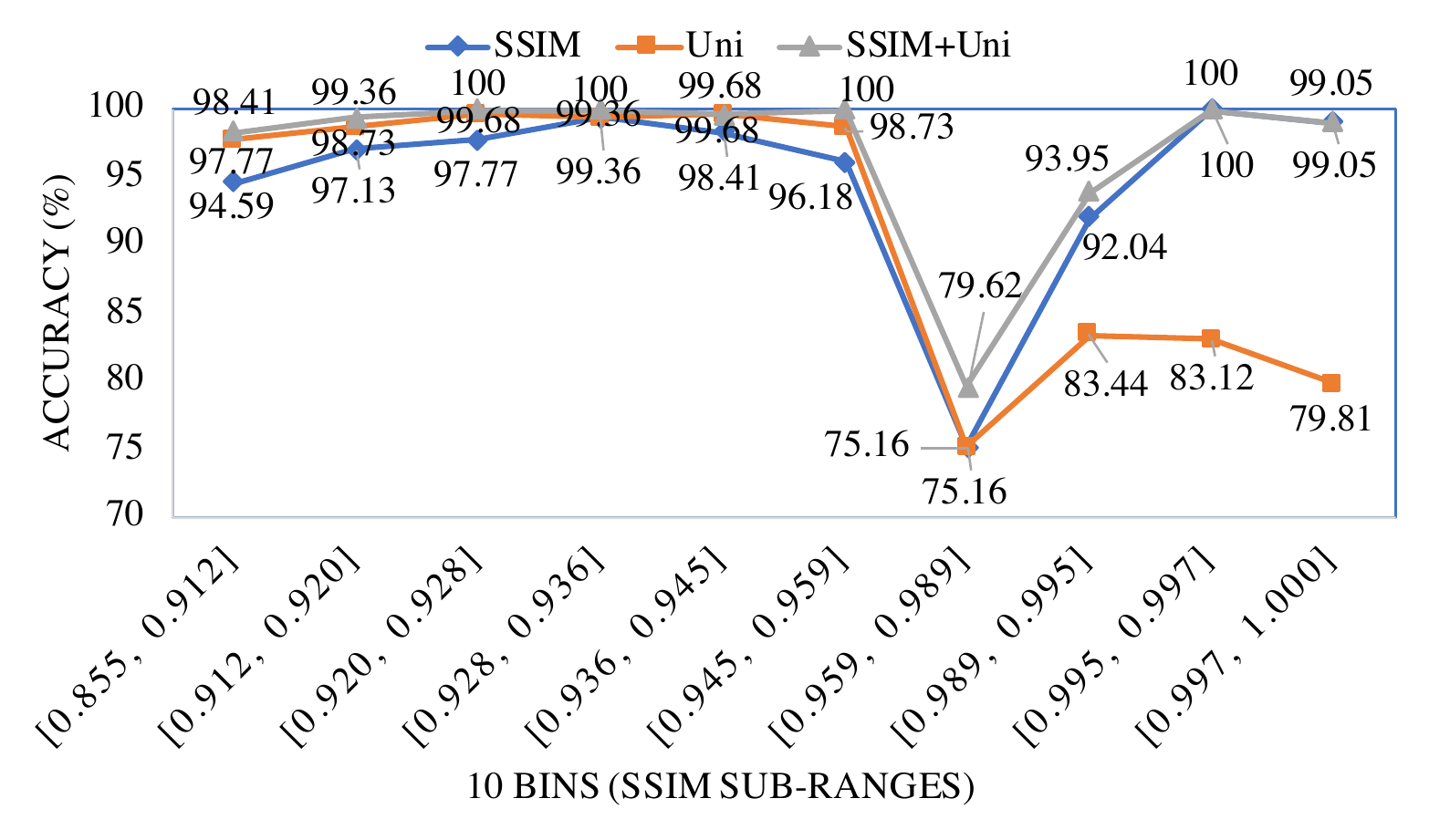}
  \caption{Accuracy for Equal Number of Samples}
  %\captionof{figure}{Accuracy for Equal Number of Samples}
  \label{fig:errorsample}
\end{figure}

\subsubsection{Error Samples in HRG}
HRG contains 945 samples including 923 homographs (97.67\%) and 22 non-homographs (2.33\%) (see Table~\ref{table:lowhigh}). We extract the error samples correctly predicted by $S_{P}$ but incorrectly predicted by $U_{P}$. 160 error samples were found in which $SU_{P}$ can correctly detect a dominant number (156 samples (97.5\%)). The 160 error samples include 155 homographs (16.79\% of homographs) and 5 non-homographs (22.73\% of non-homographs). It is reasonable when $S_{P}$ detects the homographs with high SSIMs better than $U_{P}$ because the sample that has a high SSIM score tends to be a homograph; and the sample that has a low SSIM score tends to be a non-homograph. That is why a large portion ($155/160\approx96.88\%$) of error samples are homographs. We analyzed the remaining 5 non-homographs and found that most of them have short string lengths, have lowest SSIMs in HRG, look not related at all with the brand domain, but contain special characters that appeared in other homographs. For example, \emph{baike.com} and $palk\grave{e}.com$. $palk\grave{e}.com$ is a non-homographs but contains a special character $\grave{e}$, which often appears in a homograph rather than a non-homograph. That is why $U_{P}$ cannot detect it. Although the SSIM belongs to HRG, it is not so high (0.9899), which is just asymptotic between the 7th bin (MRG) and the 8th bin (the lowest range in HRG). For such abnormal samples (non-homographs with a high SSIM), $S_{P}$ detects better than $U_{P}$.

\begin{table}
\centering
\caption{Samples in Low and High Range Groups}
\begin{tabular}{l | c | c}
 & \textbf{LRG} & \textbf{HRG}\\
 \hline
Total Samples & 1,884 & 945 \\
Homographs & 39 & 923\\
Non-homographs & 1,845 & 22\\
\hline
Total Error Samples & 44 & 160\\
Error Homographs & 17 & 155\\
Error Non-homographs & 27 & 5
\end{tabular}
\label{table:lowhigh}
\end{table}

\subsection{Homographs Registered by Brands}
A brand may proactively register homographs to protect themselves. To distinguish the homographs registered by the attacker or by the brand, Whois information can be used as mentioned in~\cite{IFIPSec19}, including organization, address, registered name, expiration date, and domain lifetime. Whois cannot be fabricated, but it can be invisible for the privacy protection. 
%!TEX encoding = UTF-8 UnicodeFor example, \emph{.ac.jp} is the top-level domain used for Japanese schools, university, or educational institutions established under the Japanese School Education Law (e.g. \emph{example.ac.jp}). The Whois of \emph{.ac.jp} managed by JPRS (Japan Registry Services) does not disclose registration address, register name, and expiration date. 
However, it is rare when all the above information is hidden. Even in such the case, other additional information can be used such as Alexa ranking, blacklists, DNS (Domain Name System), IP address, etc.

%\begin{CJK}{UTF8}{min} 
%The ``semantic'' here is not just the translation from the same meaning with different languages (for example, \emph{homograph.com} and ホモグラフ.\emph{com}), 
%\end{CJK}but also the different texts with a certain correlation (for example, \emph{apple.com} and \emph{steve-jobs.com}). While the former can be tackled using language processing, the latter requires a large social context. Furthermore, how to collect such domains for the training set is also another challenge. 

\section{Conclusion}
\label{section:conclusion}
This paper enhanced the homograph classification by~\cite{IFIPSec19} using ensemble algorithms on the combination of SSIM and unigram model. The approach can improve 1.81\% of accuracy and reduce 2.15\% of false-positive rate. We conducted an empirical error analysis and found that the SSIM cannot beat the unigram when the score is lower than 0.959, performs equally to the unigram when the score lies within $[0.959, 0.989]$, and is much better than the unigram when the score is higher than 0.989. The combination approach outperforms both the approaches in all the cases. The unigram detects low-SSIM homograph better than the SSIM, but the SSIM detects the high-SSIM non-homographs better than the unigram.  Future work investigates to address semantic homographs.

\end{document}